\documentclass[11pt]{article}
\newcommand{\be}{\begin{equation}}
\newcommand{\ee}{\end{equation}}
\newcommand{\bea}{\begin{eqnarray}}
\newcommand{\eea}{\end{eqnarray}}

\newcommand{\sn}{{\rm sn}}
\newcommand{\cn}{{\rm cn}}
\newcommand{\dn}{{\rm dn}}

\newcommand{\sss}{{\vspace{.2in}}}

\begin{document}
~\hfill{\footnotesize SUNYB/05-12, IOPB/2005-06}
\vspace{.5in}
\begin{center}
{\LARGE {\bf PT-Invariant Periodic Potentials\\ 
with a Finite 
Number of Band Gaps}}
\end{center}
\vspace{.3in}
\begin{center}
{\Large{\bf  \mbox{Avinash Khare}
 }}\\
\noindent
{\large Institute of Physics, Sachivalaya Marg, Bhubaneswar 751005, India}
\end{center}
\begin{center}
{\Large{\bf  \mbox{Uday Sukhatme}
 }}\\
\noindent
{\large Department of Physics, State University of New York at Buffalo, Buffalo, NY 14260, U.S.A. }\\
\end{center}
\vspace{.9in}
{\bf {Abstract:}}  
We obtain the band
edge eigenstates and the mid-band states for the complex, PT-invariant generalized associated 
Lam\'e potentials $V^{PT}(x)=-a(a+1)m\,\sn^2(y,m)-b(b+1)m\,{\sn^2 (y+K(m),m)}
-f(f+1)m\,{\sn^2 (y+K(m)+iK'(m),m)}-g(g+1)m\,{\sn^2 (y+iK'(m),m)}$, 
where $y \equiv ix+\beta$, and there are four parameters $a,b,f,g$. 
This work is a substantial generalization of previous work with the associated
Lam\'e potentials $V(x)=a(a+1)m\sn^2(x,m)+b(b+1)m{\sn^2 (x+K(m),m)}$
and their corresponding PT-invariant 
counterparts $V^{PT}(x)=-V(ix+\beta)$, 
both of which involving just two parameters $a,b$.  
We show that for many integer values of $a,b,f,g$, 
the PT-invariant potentials $V^{PT}(x)$ are periodic 
problems with a finite number of band gaps. Further, using
supersymmetry, we  
 construct several additional, new, complex, PT-invariant, periodic potentials with a
finite number of band gaps.  
We also point out the intimate connection between the above 
generalized associated 
Lam\'e potential problem and Heun's differential equation.

\newpage

\section{Introduction} 
In the past few years, Bender and others \cite{bed,oth} 
have looked at several complex potentials
with PT-symmetry and have shown that the energy eigenvalues 
are real when PT-symmetry is unbroken, whereas they occur in complex
conjugate pairs when PT-symmetry is spontaneously broken. 
However, there have been relatively few papers discussing
periodic potentials with PT-symmetry \cite{bdm,ks5}. 
Recently, we \cite{ks5} have constructed several new classes of analytically
solvable, complex, PT-invariant, 
periodic potentials with the special feature that they 
possess just a finite number of band gaps.
The purpose of this paper is to substantially increase this 
list of solvable periodic potentials. 

A few years ago, we obtained the band edges of
the associated Lam\'e (AL) potentials \cite{ks1}
\bea\label{1}
V(x)&&=a(a+1)m\sn^2(x,m)+b(b+1)m{\sn^2 (x+K(m),m)} \nonumber \\
&&=a(a+1)m\sn^2(x,m)+b(b+1)m\frac{\cn^2 (x,m)}{\dn^2(x,m)}\,.
\eea 
Here, $\sn \,(x,m)$, $\cn \,(x,m)$, $\dn \,(x,m)$ are Jacobi elliptic 
functions with elliptic
modulus parameter $m$ $( 0\leq m \leq 1)$. They are doubly periodic
functions with periods $[4K(m), i2K'(m)]$, $[4K(m), 2K(m)+i2K'(m)]$,
$[2K(m), i4K'(m)]$ respectively \cite{abr}, where 
$K(m) \equiv \int_0^{\pi/2} d\theta [1-m\sin^2 \theta]^{-1/2}$ 
denotes the complete elliptic integral of the first kind, and
$K'(m)\equiv K(1-m)$. For simplicity, from now on, we will not
explicitly display the modulus parameter $m$ as an argument of Jacobi
elliptic functions. 
It was shown that the AL
potentials with integral values of $a,b$ are periodic 
potentials with a finite number of 
band gaps \cite{ks2}. We also constructed and studied the PT-invariant potentials 
$V^{PT}(x) \equiv -V(ix + \beta)$ obtained from the AL potentials via the 
anti-isospectral transformation of variables $x \rightarrow ix+\beta$ \cite{ks5}.

In this paper, we make a substantial generalization of our previous work. We consider the four parameter family of generalized associated Lam\'e (GAL) potentials
\bea\label{1a}
V(x)&&=a(a+1)m\sn^2(x,m)+b(b+1)m{\sn^2 (x+K(m),m)} \nonumber \\
&&~~~~~~~~+f(f+1)m {\sn^2 (x+K(m)+iK'(m),m)}+g(g+1)m{\sn^2 (x+iK'(m),m)} \nonumber \\
&&= a(a+1)m\sn^2(x,m)+b(b+1)m\frac{\cn^2 (x,m)}{\dn^2 (x,m)} 
+f(f+1) \frac{\dn^2 (x,m)}{\cn^2(x,m)} 
+g(g+1)\frac{1}{\sn^2 (x,m)}~. \nonumber \\
\eea
In contrast to the AL potentials of eq. (\ref{1}) where there are two 
parameters $a,b$ and the two terms correspond to real translations of the 
independent variable $x$ by $0$ and $K(m)$, the GAL potentials of 
eq. (\ref{1a}) have four parameters $a,b,f,g$ and the four terms correspond to 
complex translations of the independent variable $x$ by $0, K(m), K(m)+iK'(m), 
iK'(m)$.  Although the GAL potentials are real, they do have singularities on 
the real axis coming from the zeros of the Jacobi elliptic functions 
$\sn \,(x)$ 
and $\cn \,(x)$ in the last two terms. Consequently, we will focus our 
attention on the PT-invariant versions of the GAL potentials, 
which are given by 
\bea\label{2}
V^{PT}(x)&&=-a(a+1)m\sn^2(y,m)-b(b+1)m{\sn^2 (y+K(m),m)}
 \nonumber \\
&&~~~~~~~~-f(f+1)m {\sn^2 (y+K(m)+iK'(m),m)}-g(g+1)m{\sn^2 (y+iK'(m),m)} \nonumber \\
&&= -a(a+1)m\sn^2(y,m)-b(b+1)m\frac{\cn^2 (y,m)}{\dn^2 (y,m)} 
-f(f+1) \frac{\dn^2 (y,m)}{\cn^2(y,m)} 
-g(g+1)\frac{1}{\sn^2 (y,m)} \nonumber \\
&&\equiv [a(a+1),b(b+1),f(f+1),g(g+1)]\,, 
\eea 
where 
\be\label{2a}
y=ix+\beta\,,   
\ee
with $\beta$ being an arbitrary constant. 
We shall frequently use the notation $[a(a+1),b(b+1),f(f+1),g(g+1)]$ to denote $V^{PT}(x)$. 
In this notation, PT-invariant ordinary Lam\'e potentials are denoted by
$[a(a+1),0,0,0]$, and PT-invariant AL potentials are denoted by
$[a(a+1),b(b+1),0,0]$. Here, the arbitrary constant $\beta$ is chosen
so as to avoid the singularities of the Jacobi elliptic functions on
the real axis.
We show that several of these periodic potentials for specific
integer values of $a,b,f,g$ have a finite number of band gaps. 
Looking at the symmetry of
these potentials, we are in fact tempted to conjecture that many (and perhaps all) of these 
potentials with integral values of the parameters $a,b,f,g$ also 
have a finite number of band gaps.
It would be nice if this conjecture could be proved. 

In addition, we also discover a huge class of mid-band states when
at least one of the parameters $a,b,f,g$ is half integral.
As a special case, we find some new mid-band eigenstates of the
associated Lam\'e potentials. Further, we show that the Schr\"odinger
equation for the generalized AL
potential is intimately connected with the celebrated 
Heun's differential equation \cite{ren}.
In fact, using the exact solutions obtained in this paper, one  can 
immediately obtain the corresponding solutions of Heun's equation.
In another related paper \cite{ks7}, we use this connection and
discover a wide class of new quasi-periodic solutions of Heun's
equation.

Finally, using the exact eigenstates of the GAL potentials
(\ref{2}) and the machinery of supersymmetric quantum mechanics 
\cite{cks}, we construct several more potentials with finite
band-gaps. There is one important point involved here using which we
are in fact able to construct many more supersymmetric partner
potentials corresponding to a given potential. 
The key point to note is    
that normally, in supersymmetric quantum mechanics \cite{cks}, 
given a potential $V_{-}(x)$, 
the ground state wave function $\psi_0(x)$ is used to construct the 
superpotential $W(x) = -\psi_0'(x)/\psi_0(x)$, 
which then yields the supersymmetric (SUSY) 
partner potential $V_{+}(x)=W^2+W'$. If one 
uses any excited state wave function 
$\psi(x)$ of $V_{-}(x)$ to construct a superpotential $W(x)$, then the original 
potential $V_{-}(x)$ is recovered correctly
(by construction), but 
the corresponding partner potential $V_{+}(x)$ turns
out to be singular on the real $x$-axis due to the zeros of the excited state 
wave function $\psi(x)$. However, as has been noticed recently \cite{rl},
if we consider PT-symmetric complex potentials, then the singularity is not
on the real axis. Besides, as we have stressed previously \cite{ks5,ks6}, 
in the case of doubly periodic potentials 
composed of Jacobi elliptic functions, both $V(x)$ and $V^{PT}(x)$ can
be simultaneously periodic even though their periods are different.
In this way, by starting from the analytically solvable Lam\'e and 
associated Lam\'e potentials and using the  excited state band edges
of the corresponding PT-symmetric potentials, we discover a wide range
of new, analytically solvable, complex PT-invariant periodic potentials 
with a finite number of band gaps. As 
an illustration, we discuss a few of
these potentials in detail. 

The plan of the paper is the following. 
In Sec. 2 
we discuss the PT-invariant GAL potentials (\ref{2}) 
in some detail
and obtain band edges as well as mid-band states of several of these
potentials. As a byproduct we also obtain some new solutions of the
AL potentials (which we had missed in earlier work
\cite{ks1,ks2}). Further, we show that the class of potentials
$[a(a+1),0,0,g(g+1)]$ have finite number of band-gaps in case $a,g$
are integers. 
In Sec. 3 we start from the energy eigenstates obtained in
Sec. 2 and using both the ground state as well as excited state wave
functions,
obtain new periodic PT-invariant potentials with a 
finite number of band gaps. 
In Sec. 4 we briefly discuss the connection between the
solutions of the potentials (\ref{2}) and Heun's differential 
equation.

\section{Solutions for the Generalized Associated Lam\'e (GAL) Potentials}
A few years ago, we obtained analytic solutions of the associated
Lam\'e potentials (\ref{1}) \cite {ks1,ks2} and showed that when
$a,b$ are integers, then the resulting potentials all had a
finite number of band gaps.
The purpose of this
section is to show that the complex
PT-invariant GAL potentials as given by eq. (\ref{2}) 
are also quasi-exactly solvable. In particular, we show that the band
edges or mid-band states of these problems can be obtained
depending on whether $a+b+f+g$ (or $a-b-f-g$) is an integer or an arbitrary
non-integer number. 
It should be noted that we are considering
PT-invariant potentials (\ref{2}), since the corresponding real 
potentials (\ref{1a}) are singular on the real axis.

It may be worthwhile explaining the underlying 
basic idea here, even though it has been well
established by us before \cite{ks5}. Note 
that if $\psi(x)$ is a solution 
of the Schr\"odinger equation for the real potential 
$V(x)$ with energy $E$, 
then $\psi(ix+\beta)$ is a solution of the Schr\"odinger equation 
for the complex potential 
$-V(ix+\beta)$ with energy $-E$, where $\beta$ is an arbitrary nonzero  
constant. The new potential $-V(ix+\beta)$, generated by the anti-isospectral 
transformation $x \rightarrow ix+\beta$ \cite{kuw}, 
is clearly PT-symmetric and will be denoted 
by $V^{PT}(x)$. Further, if $\psi(x)$ and $\psi(ix+\beta)$ 
satisfy appropriate boundary 
conditions, they are eigenfunctions of $V(x)$ and $V^{PT}(x)$ respectively. The 
ordering of energy levels for $V^{PT}(x)$ is the opposite of the ordering of 
energy levels for $V(x)$. 

In this paper, our main focus is on the Schr\"odinger equation ($\hbar=2m=1$)
\be\label{3.2}
-\frac{d^2}{dx^2}\psi(x)+V^{PT}(x)\psi(x)=E\psi(x)\,,
\ee
where $V^{PT}(x)$ is the potential given by eq. (\ref{2}). 
Eq. (\ref{3.2}) is called the generalized associated
Lam\'e equation, and we are seeking its eigenstates and mid-band states.

\subsection{\bf Symmetries}

At this stage, it is worth pointing out the symmetries of 
the PT-invariant GAL potential (\ref{2}) 
and hence the corresponding Schr\"odinger equation (\ref{3.2}).

\begin{enumerate}

\item The potential (\ref{2}) and hence the 
Schr\"odinger eq. (\ref{3.2}) 
remains unchanged when any one (or more) 
of the four parameters $a,b,f,g$
change to $-a-1,-b-1,-f-1,-g-1$ respectively. 

\item Under the translation $y \rightarrow y+K(m)$, the GAL
potential $[a(a+1),b(b+1),f(f+1),g(g+1)]$ goes to the potential
$[b(b+1),a(a+1),g(g+1),f(f+1)]$. 
Hence, both GAL potentials
must have the same energy eigenvalues and the corresponding
energy eigenfunctions are simply related:
$y \rightarrow y+K(m)$, i.e.
\be\label{3a} 
E^{PT}(b,a,g,f;m) = E^{PT}(a,b,f,g;m)\,,~~\psi(y,b,a,g,f;m) \propto
\psi(y+K(m),a,b,f,g;m)\,.
\ee

\item Similarly, 
by considering the translations $y \rightarrow y+K(m)+iK'(m)$, 
and $y \rightarrow y+iK'(m)$, it is easy to show that
\be\label{1b}
E^{PT}(f,g,a,b;m) = E^{PT}(a,b,f,g;m)\,,~~\psi(y,f,g,a,b;m) \propto
\psi(y+K(m)+iK'(m),a,b,f,g;m)\,.
\ee
\be\label{1c}
E^{PT}(g,f,b,a;m) = E^{PT}(a,b,f,g;m)\,,~~\psi(y,g,f,b,a;m) \propto
\psi(y+iK'(m),a,b,f,g;m)\,.
\ee

\end{enumerate}
Thus, once we obtain the eigenvalues and eigenfunctions of a given
GAL potential $[a(a+1),b(b+1),f(f+1),g(g+1)]$, then
we immediately know the eigenvalues and eigenfunctions 
 of three other potentials: 
$[b(b+1),a(a+1),g(g+1),f(f+1)]$, $[f(f+1),g(g+1),a(a+1),b(b+1)]$ and
$[g(g+1),f(f+1),b(b+1),a(a+1)]$. Therefore, it suffices to 
present results for only one of the four potentials.

\subsection{\bf Duality Relations}

We shall now derive some remarkable relations
relating the quasi-exactly solvable 
eigenvalues and eigenfunctions (corresponding either to
the band edges or mid-band states) of two GAL 
potentials at two different 
values $m$ and $1-m$ of
the modulus parameter. 

To that purpose we start from the Schr\"odinger eq. (\ref{3.2}) for
the PT-invariant GAL potential (\ref{2}). On using the relations
\cite{abr} 
\bea\label{3.2a}
&&\sqrt{m}\,\sn(y,m)=-\dn\,[iy+K'(m)+iK(m),1-m]\,, \nonumber \\
&&\dn(y,m)=\sqrt{1-m}\, \sn\,[iy+K'(m)+iK(m),1-m]\,, \nonumber \\
&&\sqrt{m}\,\cn(y,m)=i\sqrt{1-m}\, \cn\,[iy+K'(m)+iK(m),1-m]\,,
\eea
and defining a new variable $w=iy+K'(m)+iK(m)$, the Schr\"odinger 
eq. (\ref{3.2}) takes the form
\bea\label{3.2b}
&&\-\psi''(w)-[a(a+1)(1-m)\sn^2(w,1-m)+g(g+1)(1-m)\frac{\cn^2(w,1-m)}
{\dn^2(w,1-m)} 
f(f+1)\frac{\dn^2(w,1-m)}{\cn^2(w,1-m)} \nonumber \\
&&+b(b+1)\frac{!}{\sn^2(w,1-m)}]
\psi(w)=-[a(a+1)+b(b+1)+f(f+1)+g(g+1)+E]\psi(w)\,.
\eea
On comparing eqs. (\ref{3.2}) and (\ref{3.2b}) we then have the
remarkable relations
\bea\label{3.2c}
&&E^{PT}(a,b,f,g,m)=-[a(a+1)+b(b+1)+f(f+1)+g(g+1)]
-E^{PT}(a,g,f,b,1-m)\,, \nonumber \\
&&\psi(y,m) \propto \psi(iy+K'(m)+iK(m),1-m)\,,
\eea
which is valid for the QES states corresponding to either the band
edges or mid-band states. Note that here, $a,b,f,g$ can be arbitrary
(real) numbers and are not restricted to integer values. 
This is a very powerful relation which has several interesting
consequences. One immediate
important consequence of 
eq. (\ref{3.2c}) is that
for arbitrary integer values of $a,g$, 
the potential $[a(a+1),0,0,g(g+1)]$ has only a finite number of band-gaps. This happens because, for 
$f=g=0$, one has
\be\label{3.2d}
E^{PT}(a,b,0,0,m)=-[a(a+1)+b(b+1)]-E^{PT}(a,0,0,g=b,1-m)\,,
\ee
so that both the potentials must have the same number of band-edges and
band-gaps and we have already proved \cite{ks2} that the AL potentials
have finite number of band gaps in case $a,b$ are integers. 

\subsection{\bf QES Solutions}

Let us now seek solutions of the Schr\"odinger eq. (\ref{3.2}) for the
PT-invariant GAL potential (\ref{2}).
On making the ansatz
\be\label{3.3}
\psi(x)=\dn^{-b}(y) \sn^{-g}(y) \cn^{-f}(y) \phi
(y)\,,~~y=ix+\beta\,,
\ee
it is easily shown that $\phi$ satisfies the equation
\be\label{3.4}
\phi''(y)+2[mb\frac{\sn(y)\cn(y)}{\dn(y)}-g\frac{\cn(y)\dn(y)}
{\sn(y)}+f\frac{\dn(y)\sn(y)}{\cn(y)}]\phi'(y) 
+[Qm\sn^2(y)-R]\phi(y)=0\,,
\ee
where
\be\label{3.5}
Q=(b+g+f)(b+g+f-1)-a(a+1)\,,~~R=E+(f+g)^2+m(g+b)^2\,.
\ee
It is well known \cite{bre} 
that this is a quasi-exactly solvable (QES) problem.  
We shall now systematically consider solutions of eq. (\ref{3.4}) 
for several special cases and then finally consider 
the most general case.

\subsection{\bf $b=f=g=0$}

The simplest possibility is when three out of the
four parameters $a,b,f,g$ are zero. For example, when $b=f=g=0$, then
the problem reduces to the PT-invariant version of the  
well studied Lam\'e potential problem. We might add here that, instead of $a$, if any one
of the other parameters $b,f,g$ is nonzero, one still has a potential which is strictly
isospectral to the PT-invariant Lam\'e potential.
It may be noted that while the Lam\'e potential is a periodic
potential with (real) period $2K(m)$, the PT-invariant Lam\'e potential
has real period $2K'(m)$. Further, the band edge eigenvalues,
eigenfunctions and the discriminant $\Delta$ of $V^{PT}(x)$ are
related to those of Lam\'e potential by \cite{ks5}
\bea\label{6}
&&E_j^{PT}(m) =
-E_{2a-j}(m)\,,~~~\psi_j^{PT}(x,m)~\propto~\psi_{2a-j}(ix+\beta,m)~,
~~~j=0,1,2,...\,
,2a\, \nonumber \\
&&\Delta^{PT} (E,m)=\Delta[E+a(a+1),1-m]\,. 
\eea  

{}From eq. (\ref{3.2c}), it follows that the PT-invariant Lam\'e 
band-edge eigenvalues and
eigenfunctions, for integral $a$ 
satisfy the remarkable relations ($j=0,1,2,...2a$)
\be\label{55a}
E^{PT}_{j}(m)=-a(a+1)-E^{PT}_{2a-j}(m)\,,~~\psi_{j}(y,m) 
\propto \psi_{2a-j}(iy+K'(m)+iK(m),1-m)\,.
\ee
We would like to add here that even the mid-band states satisfy (for
half-integral $a$)
relations analogous to (\ref{55a}):
\be
E_{j}(m)=a(a+1)-E_{a-1/2-j}(m)\,,~~\psi_{j}(y,m) 
\propto \psi_{a-1/2-j}(iy+K'(m)+iK(m),1-m)\,,
\ee
where $j=0,1,2,...,a-1/2$ .
Note the remarkable fact that for any integer $a$, all  bands and band gaps 
exchange their role as one goes from the Lam\'e potential 
to its PT-invariant version $V^{PT}(x)$ \cite{ks5}.

The next simple possibility is when two of the four parameters $a,b,f,g$  are
zero. Here there are three distinct possibilities which we discuss one
by one.

\subsection{\bf $f=g=0$}

In this case the problem reduces to the PT-invariant AL potential
which we have already discussed at great length \cite{ks1,ks2}. 
Note that if either $a$ or $b$ is
zero (or -1), then this 
potential reduces to the PT-invariant of the Lam\'e potential. 
As previously shown by us \cite{ks2}, for arbitrary integral values of $a$ and
$b$, AL potentials 
are exactly solvable problems 
with finite number of
band-gaps for which
one can write down the form of all the
band edge eigenfunctions, as we do below. We note here that
when $a > b$ are both integers, then there are precisely $a$
bound bands (some of which are unusual in that both the band edges are
of the same period), same ($a$) number of band gaps and all
the $2a+1$ band edges are analytically known beyond which there is a
continuum band extending to $E= \infty$. Note that if $b > a$,
then also there are $b$ bound bands and $b$ band gaps and 
the corresponding eigenfunctions are simply obtained from the
$a>b$ case by the transformation $x \rightarrow x+K(m)$ while the $a=b$
case essentially corresponds to the Lam\'e potential $[a(a+1),0,0,0]$.
Without any loss of generality, we shall only consider AL
potentials with $a > b$. 

The form of the $2a+1$ band edge eigenfunctions of the AL potential
depends on whether $a-b$ is an odd or an even integer. For example, when
$b=a-2p-1\, (p=0,1,2,...$), then there are: 

$p$ eigenstates of the form
$\sn(x)\cn(x)\dn(x) F_{p-1} [\sn^2(x)]$, 

$p+1$ eigenstates of the form
$\dn^{a-2p}(x) F_{p} [\sn^2(x)]$, 

$a-p$ eigenstates of the form
$\cn(x)\dn^{2p+1-a}(x) F_{a-p-1} [\sn^2(x)]$,

$a-p$ eigenstates 
of the form
$\sn(x)(\dn^{2p+1-a}(x) F_{a-p-1} [\sn^2(x)]$.
 
\noindent On the other hand, when $b=a-2p$, there are: 

$p$ eigenstates of the
form $\cn(x)\dn^{a-2p+1}(x) F_{p-1} [\sn^2(x)]$, 

$p$ eigenstates 
of the form
$\sn(x)\dn^{a-2p+1}(x) F_{p-1} [\sn^2(x)]$, 

$a-p$ eigenstates
of the form
$\sn(x)\cn(x)\dn^{2p-a}(x)F_{a-p-1} [\sn^2(x)]$, 

$a-p+1$ 
eigenstates of the form
$\dn^{2p-a}(x) F_{a-p} [\sn^2(x)]$.

\noindent Here $F_n [\sn^2(x)]$ denotes a polynomial in $\sn^2(x)$ of order
$n$.

We would like to re-state here that all the eigenstates of the PT-invariant version 
of the AL potentials are immediately obtained from the
known eigenfunctions of the 
associated Lam\'e problem and the ordering of energy levels  
of these is the opposite of the corresponding AL problem. Hence, this
is also an exactly solvable problem 
with a finite number ($a$) of band gaps and $2a+1$ known band edges when 
both $a,b$ are integers.

\subsection{\bf $b=f=0$}

Following our discussion for the AL case, 
without any loss of generality we assume here that $a > g$.
In this case, one obtains $n+1$ QES solutions when $a+g=n$
(or $g-a=n+1$) with $n=0,1,2,...$. 
The QES solutions for $n=0,1,2,3,4$ are given in Table 1. In
particular, for any choice of $a(a+1)$, Table 1 lists the
eigenstates for various values of $g(g+1)$. The general form of these
eigenfunctions is obtained from the corresponding AL
eigenfunctions as given in Table 3 of \cite{ks1} by simply
interchanging $\dn(y)$ and $\sn(y)$.

A few remarks are in order.

\begin{enumerate}

\item Since we are considering the case ($b=f=0$), the duality relation 
(\ref{3.2c}) takes the form
\bea\label{3.2e}
&&E^{PT}(a,0,0,g;m)=-[a(a+1)+g(g+1)]-E^{PT}(a,b=g,0,0;1-m)\,,
\nonumber \\
&& \psi(y,a,0,0,g;m)
\propto \psi(iy+K'(m)+iK(m),a,b=g,0,0;1-m)\,,
\eea
Using Table 3 of ref. \cite{ks1} and this duality relation,
it is straightforward to obtain all the QES eigenstates,
thereby providing an independent check on the
results given in Table 1. Further, it follows that for arbitrary
integer values of $a$ and $g$, $[a(a+1),0,0,g(g+1)]$ is an 
exactly solvable potential problem with a finite number ($a$) of
band-gaps. From the duality relation (\ref{3.2e}), 
it follows that for integer values of $a,g$  
\be\label{3.2f}
E_j^{PT}(a,0,0,g;m)=-[a(a+1)+g(g+1)]+E_j (a,b=g,0,0;1-m)\,,
\ee
and hence the corresponding discriminants $\Delta$ are related by
\be\label{3.2g}
\Delta^{PT}(E,m;a,0,0,g)=\Delta[E+a(a+1)+g(g+1),1-m;a,b=g,0,0]\,.
\ee

\item Following the structure of the eigenfunctions of the AL
potentials as given above, it is now straightforward 
to write down the general form of the eigenfunctions for arbitrary value
of $n$. However, to obtain the corresponding eigenvalues, one needs to 
solve cubic and higher order equations.

\item Under the transformation $y \rightarrow y+iK'(m)$ followed by
the interchange of $a$ and $g$ (note $b=f=0$), the Schr\"odinger eq.
(\ref{3.2}) for the GAL potential (\ref{2}) remains unchanged. Thus
it follows that under the interchange of $a$ with $g$, the eigenvalue
spectrum must remain unaltered. Clearly, this is only possible if
either the energy eigenvalues remain unchanged under this
transformation, or if two of the eigenvalues go into each other. It is
easy to verify from Table 1 that the eigenvalues corresponding 
to the eigenfunctions of period
$2iK'(m)$ remain unaltered under
$a \rightarrow g$ while the other eigenvalues  
go into each other under this transformation.

\item Similarly, From Table 3 of \cite{ks1}, it is easy to check that for the AL
potentials (\ref{1}), 
the eigenvalues corresponding 
to the eigenfunctions of period
$2K(m)$ remain unaltered under
$a \rightarrow b$ while the other eigenvalues  
go into each other under this transformation. This happens because the
AL potentials remain unaltered under the transformation
$y \rightarrow y+K(m)$ followed by the interchange of $a$ with $b$. 

\end{enumerate}

Summarizing, we have discovered new exactly solvable potential
problems with a finite number of band gaps when $a,g$ are arbitrary
integers. In fact everything about these potentials can be derived
from previously known results for AL potentials.

\subsection{$b=g=0$}

In this case, one obtains $n+1$ QES solutions when $a+f=n$
with $n=0,1,2,...$.
The solutions for $n=0,1,2,3,4$ are given in Table 2. In
particular, for any choice of $a(a+1)$, Table 2 lists the
eigenstates for various values of $f(f+1)$. The general form of these
eigenfunctions is simply obtained from the corresponding AL
eigenfunctions as given in Table 2 of \cite{ks1} by 
interchanging $\dn(y)$ and $\cn(y)$.

Some comments are in order at this stage.

\begin{enumerate}

\item The form of eigenfunctions for arbitrary value of $n$ is easily
written down following the structure of the AL eigenfunctions given in
the last section. 

\item From eq. (\ref{3.2c}) it follows that the potential (\ref{2})
with $b=g=0$ is a self-dual potential, satisfying
\be
E^{PT}_{j_1}(a,f,m)=-[a(a+1)+f(f+1)]-E^{PT}_{j_2}(a,f,1-m)\,.
\ee
Using Table 2, it is easily checked that indeed this is true, 
for any values of $a,f$. In particular, whereas
$\delta_{5},\delta_{8}$ are invariant under $m \rightarrow 1-m$, 
$\delta_6 \leftrightarrow \delta_7$ under the same transformation.

\item Under the transformation $y \rightarrow y+K(m)+iK'(m)$ followed by
the interchange of $a$ and $f$ (note $b=g=0$), the Schr\"odinger eq.
(\ref{3.2}) for the GAL potential (\ref{2}) remains unchanged. Thus
it follows that under the interchange of $a$ with $f$, the eigenvalue
spectrum must remain unaltered. Clearly, this is only possible if
either the energy eigenvalues remain unchanged under this
transformation, or if two of the eigenvalues go into each other. It is
easy to verify from Table 2 that the eigenvalues corresponding 
to the eigenfunctions of period
$2K(m)+2iK'(m)$ remain unaltered under
$a \rightarrow f$ while the other eigenvalues  
go into each other under this transformation. In particular, while
$\delta_5,\delta_8$ are invariant under $a \leftrightarrow f$,
$\delta_6 \leftrightarrow \delta_7$ under the same transformation.

\end{enumerate}

\subsection{$f=0$}

Let us consider the case when only one out of the four parameters $a,b,f,g$ is zero. As an 
illustration, we discuss the case $f=0$. In fact, as described below, once
we know the eigenstates of this problem, the eigenstates of the other
three problems corresponding to either $b$ or $g$ or $a$ equal to zero
are immediately obtainable, since the four potentials are related by translations of the independent variable.

For the case $f=0$, one obtains $\frac{n+2}{2} (\frac{n+1}{2})$ QES solutions 
when $n$ is even (odd). Here $a+b+g=n$ 
with $n=0,1,2,...$.
The QES solutions for $n=0,1,2,3$ are given in Table 3. In
particular, for any choice of $a(a+1)$, Table 3 lists the
eigenstates for various values of $(b+g)(b+g+1)$. 

Some remarks are appropriate.

\begin{enumerate}

\item By looking at the structure of the QES eigenfunctions in Table
3, it is easy to write down the nature of eigenfunctions
for the general case.

\item From Table 3, it is easily checked that the duality relation 
\be\label{3.2h}
E^{PT}(a,b,g,m)=-[a(a+1)+b(b+1)+g(g+1)]-E^{PT}(a,g,b,1-m)\,.
\ee
is indeed satisfied. In particular, both
$\delta_9,\delta_{10}$ are invariant under $b \leftrightarrow g$
followed by $m \rightarrow 1-m$.

\item Under the transformation $y \rightarrow y+K(m)$ followed by
the interchange of $a$ and $b$, and replacing $g$ by $f$, the Schr\"odinger eq.
(\ref{3.2}) for the GAL potential (\ref{2}) with $f=0$ goes
over to the Schr\"odinger equation for the GAL potential
(\ref{2}) with $g=0$. Hence, under the interchange of $a$ and $b$
and replacing $g$ by $f$, all the energy eigenvalues of the potential
(\ref{2}) with $f=0$ must go over into those of (\ref{2}) with $g=0$,
while the corresponding eigenfunctions are simply obtained from Table
3 by replacing $y$ by $y+K(m)$.

\item Using similar reasoning it also follows that under the interchange
of $a$ with $g$ and replacing $b$ by $f$, all the energy eigenvalues
of the GAL potential (\ref{2}) with $f=0$ go over to
those of potential (\ref{2}) with $b=0$ while the corresponding
eigenfunctions are obtained from Table 3 by replacing $y$ by 
$y+iK'(m)$. 
And finally,
under the interchange
of $b$ with $g$ and replacing $a$ by $f$, all the energy eigenvalues
of the GAL potential (\ref{2}) with $f=0$ go over to
those of potential (\ref{2}) with $a=0$, while the corresponding
eigenfunctions are easily obtained from Table 3 by replacing $y$ by
$y+K(m)+iK'(m)$. 

\end{enumerate}

\subsection{The General Case: $a,~b,~f,~g$ All Nonzero}

Finally, let us discuss the most general case when all the four
parameters are nonzero. 
In this case one obtains $n+1$ solutions when $a+b+f+g=2n$ with
$n=0,1,2,...$. The QES solutions for $n=0,1$ are given in Table 4.

\begin{enumerate}

\item It is easy to see that in the general case, the eigenfunction is
of the form
\be
\psi = \sn^{-g}(y)\cn^{-f}(y)\dn^{-b}(y) \sum_{k=0}^{n} A_k \sn^{2k}
(y)\,,
\ee
while the corresponding eigenvalues are solutions of a $n+1$'th order
equation.

\item It can be checked from Table 4 that $\delta_{11}$ is
invariant under $b \leftrightarrow g$ followed by $m \rightarrow 1-m$.

\item The GAL potential (\ref{2}) and hence the
corresponding Schr\"odinger eq. (\ref{3.2}) is invariant under
the transformation $y \rightarrow y+K(m)$ followed by the interchange
of $a$ with $b$ and $f$ with $g$. Hence, under the interchange of
$a$ with $b$ and $f$ with $g$, all the eigenvalues of the GAL system must
either remain invariant or go into each other. In fact it is easily
checked from Table 4 that all
the eigenvalues  are invariant under the
interchange of $a$ with $b$ and $f$ with $g$. Extending this argument,
in fact one finds that all the eigenvalues are also invariant under 
$a \leftrightarrow f, b \leftrightarrow g$ as well as under
$a \leftrightarrow g, b \leftrightarrow f$.

\end{enumerate} 

\subsection{Mid-Band States}

So far we have discussed the results for the PT-invariant GAL potentials,
which give eigenvalues and eigenfunctions corresponding to the
band edges. It may be noted that in all these cases, while $a,b,f,g$ 
need not be integers, either $a+b+f+g$ or $a-b-f-g$ is always
integral. We now show that when at least one of $a,b,f,g$ is 
half-integral and either $a+b+f+g$ and/or $a-b-f-g$
is an arbitrary number (being an integer is of course a very special
case here), then one can obtain doubly degenerate eigenstates
which correspond to mid-band states. In fact depending on whether we want 
$b$ or $f$ or $g$ to be half-integral (with the other two parameters being integral), 
we need to use different trial solutions. Therefore, we shall consider all three cases one
by one.

{\bf Case 1: $b$ half-integral}

We start from eq. (\ref{3.4}) and further substitute the ansatz
\be\label{3.9}
\phi(y) = [\cn(y)+i\sn(y)]^{t} Z(y)\,,
\ee
where $t$ is any real number. After lengthy but straightforward 
algebra, one can show that $Z(y)$ satisfies the equation
\bea\label{3.10}
&&Z''(y)+[2it\dn(y)+2mb\frac{\sn(y)\cn(y)}{\dn(y)}
-2g\frac{\cn(y)\dn(y)}{\sn(y)}+2f\frac{\dn(y)\sn(y)}
{\cn(y)}]Z'(y) \nonumber \\
&&+[-(R+t^2)+(Q+t^2)m\sn^2(y)-2itg\frac{\cn(y)}{\sn(y)}
+2itf(1-m)\frac{\sn(y)}{\cn(y)}\nonumber \\
&&+imt(2b+2f+2g-1)\sn(y)\cn(y)]Z(y)=0\,,
\eea 
where $R$ and $Q$ are as given by eq. (\ref{3.5}).
Not surprisingly, $Z(y)=$constant is a solution with energy
$E=-(4t^2+m)/4$ provided $f=g=0,b=1/2,a=t-1/2$ (i.e. $b+f+g=1/2$).  

One can build solutions for higher values of $b+f+g$ from here. In
particular, for $b+f+g=2M+1/2$, we consider the ansatz
($M=0,1,2,...$)
\be\label{3.11}
Z(y)=\sum_{k=0}^{M} A_k \sn^{2k}
(y)+\cn(y)\sn(y)\sum_{k=0}^{M-1} B_k \sn^{2k} (y)\,,
\ee
while if $b+f+g=2M+3/2$ then we consider the ansatz ($M=0,1,2,...$)
\be\label{3.12}
Z(y)=\cn(y)\sum_{k=0}^{M} A_k \sn^{2k}
(y)+\sn(y)\sum_{k=0}^{M} B_k \sn^{2k} (y)\,,
\ee
Substitution into eq. (\ref{3.10}) and simplification yields
analytic expressions for the
energy eigenvalues and eigenfunctions for arbitrary $M$ for 
$b=1/2$ and $b=3/2$. In particular, for $b=1/2$, we find that
\be\label{3.13}
b=1/2,f=p,f+g=N,a=t-1/2\,,~~E=-[t^2+m(g+b)^2]\,,
\ee
where both $f,g$ are nonnegative integers satisfying $f+g=N$ with
$N=0,1,2,...$.

Similarly, when $b=3/2,a=t-1/2,f=p,f+g=N$ we find that
\be\label{3.14}
E=m(2g+1)-[1+t^2+m(g+b)^2]\pm \sqrt{(2g+1)^2m^2
+4m(N+1)(f-g)+4(1-m)t^2}\,
\ee
where, $f$ and $g$ are again nonnegative integers. In all these cases, the
corresponding eigenfunctions have the form as given above in eqs.
(\ref{3.11}) and (\ref{3.12}). For small
 values of N, the explicit coefficients $A_k,B_k$ appearing in the
eigenfunction expressions can be easily written
down. For example, for $b=1/2$ and $N=1$, the 
eigenfunction is $Z(y)=A\cn(y)+B\sn(y)$ with $\frac{B}{A}=it$ in
case $f=1,g=0$ while $\frac{B}{A}=i$ in case $g=1,f=0$.

For the special case of $f=g=0$ and $t \ne 1/2$, these results
represent the generalization of results obtained by us previously
\cite{ks2} in the case of AL potential. Further, for $f=g=0,t=1/2$, 
the results obtained above match with 
the energy eigenvalue expressions obtained in ref. 
\cite{ks2} (as they should!).

Several comments can be readily made.

\begin{enumerate}

\item Since, in the variable y,
 the GAL potential (\ref{2}) has period $2K(m)$
as well as $2iK'(m)$, hence $\psi(y)$ and $\psi(y+2K(m))$ as well as
$\psi(y+2iK'(m))$ are all eigenfunctions of GAL equation
with the same eigenvalue. As a consequence,  
$\phi(y)=[\cn(y)-i\sn(y)]^{t}Z(y)$ is also the eigenfunction with
the {\it same} eigenvalue. Thus for any nonintegral $t$, each level is
doubly-degenerate. The same remark also applies to the other two
solutions (when $f$ or $g$ is half integral) discussed below.

\item There is one remarkable symmetry associated with eq.
(\ref{3.10}). In particular, notice that this equation is invariant under 
$t \rightarrow -t$ followed by $i \rightarrow -i$ (where
$i=\sqrt{-1}$). But under this transformation, the ansatz (\ref{3.9})
becomes
\be\label{3.9aa}
\phi(y)=[\cn(y,m)-i\sn(y,m)]^{-t}\,,
\ee
Hence it follows that the energy eigenvalues must be independent of
sign of $t$, i.e. they must be a function of $t^2$.
Similar remarks also apply in the other two cases discussed below
(i.e. when $f,g$ are half-integral).

\item For integral $t$, both $a,b$ are half integral and these
solutions reduce to those discussed in the last section and in that
case they correspond to QES band edge eigenstates.

\item Here we have obtained solutions $\psi(y)$ in which $a =t-1/2,f=p,g=N-p$
and $b=1/2$ or $3/2$. In view of the symmetries of the GAL
potentials, we then also have solutions
$\psi(y+K(m))$ with the same energy 
in case $b=t-1/2, g=p, f=N-p$ and $a$ is either $1/2$ or
$3/2$. Similarly we have solutions $\psi(y+K(m)+iK'(m))$ with the same
energy in case $f=t-1/2,a=p,b=N-p$ and $g=1/2$ or $3/2$. Further, we 
also have solutions $\psi(y+iK'(m))$ with the same energy in case 
$g=t-1/2, a=N-p, b=p$ and $f=1/2$ or $3/2$.

\end{enumerate}

{\bf Case 2: $f$ half-integral}

We start from eq. (\ref{3.4}) and further substitute the ansatz
\be\label{3.15}
\phi(y) = [\dn(y)+ik\sn(y)]^{t} Z(y)\,,
\ee
where $t$ is any real number and $k = \sqrt{m}$. 
After some lengthy but straightforward 
algebra, one finds that $Z(y)$ satisfies the equation
\bea\label{3.16}
&&Z''(y)+[2ikt\cn(y)+2mb\frac{\sn(y)\cn(y)}{\dn(y)}
-2g\frac{\cn(y)\dn(y)}{\sn(y)}+2f\frac{\dn(y)\sn(y)}
{\cn(y)}]Z'(y) \nonumber \\
&&+[-(R+m t^2)+(Q+t^2)m\sn^2(y)-2itkg\frac{\dn(y)}{\sn(y)}
-2iktb(1-m)\frac{\sn(y)}{\dn(y)}\nonumber \\
&&+ikt(2b+2f+2g-1)\sn(y)\dn(y)]Z(y)=0\,,
\eea 
where $R$ and $Q$ are as given by eq. (\ref{3.5}).
Not surprisingly, $Z(y)=$constant is a solution with energy
$E=-(4mt^2+1)/4$ provided $b=g=0,f=1/2,a=t-1/2$ (i.e. $b+f+g=1/2$).  

One can build solutions for higher values of $b+f+g$ from here. In
particular, in case $b+f+g=2M+1/2$, we consider the ansatz
($M=0,1,2,...$)
\be\label{3.17}
Z(y)=\sum_{k=0}^{M} A_k \sn^{2k}
(y)+\sn(y)\dn(y)\sum_{k=0}^{M-1} B_k \sn^{2k} (y)\,,
\ee
while if $b+f+g=2M+3/2$ then we consider the ansatz ($M=0,1,2,...$)
\be\label{3.18}
Z(y)=\dn(y)\sum_{k=0}^{M} A_k \sn^{2k}
(y)+\sn(y)\sum_{k=0}^{M} B_k \sn^{2k} (y)\,,
\ee
On substituting this ansatz in eq. (\ref{3.16}) and making algebraic
simplifications, we obtain analytic expressions for the
energy eigenvalues and eigenfunctions for arbitrary $M$ for
$f=1/2$ and $f=3/2$. In particular, for $f=1/2$, we find that
\be\label{3.19}
f=1/2\,,~ b+g=N\,,~ a=t-1/2\,,~~E=-[mt^2+(g+f)^2]\,,
\ee
where both $b,g$ are nonnegative integers satisfying $b+g=N$ with
$N=0,1,2,...$.

Similarly, when $f=3/2,a=t-1/2,b+g=N$ we find that
\be\label{3.20}
E=(2g+1)-[(1+t^2)m+(g+f)^2] \pm \sqrt{(2g+1)^2
+4m(N+1)(f-g)-4m(1-m)t^2}\,
\ee
where, $b$ and $g$ are again nonnegative integers. In all these cases, the
corresponding eigenfunctions have the form as given above in eqs.
(\ref{3.17}) and (\ref{3.18}). 
For small values of N, the explicit coefficients $A_k,B_k$ in the
eigenfunction expressions can be easily written
down. Further, as in the half-integral $b$ case, one can write down
three more solutions with the same energy. 

{\bf Case 3: $g$ half-integral}

We start from eq. (\ref{3.4}) and further substitute the ansatz
\be\label{3.21a}
\phi(y) = [\dn(y)+k\cn(y)]^{t} Z(y)\,,
\ee
where $t$ is any real number. After algebraic simplification,
it is easy to show that $Z(y)$ satisfies the equation
\bea\label{3.21}
&&Z''(y)+[-2kt\sn(y)+2mb\frac{\sn(y)\cn(y)}{\dn(y)}
-2g\frac{\cn(y)\dn(y)}{\sn(y)}+2f\frac{\dn(y)\sn(y)}
{\cn(y)}]Z'(y) \nonumber \\
&&+[-R+(Q+t^2)m\sn^2(y)-2ktb\frac{cn(y)}{\dn(y)}
-2ktf\frac{\dn(y)}{\cn(y)}\nonumber \\
&&+kt(2b+2f+2g-1)\cn(y)\dn(y)]Z(y)=0\,,
\eea 
where $R$ and $Q$ are as given by eq. (\ref{3.5}).
Not surprisingly, $Z(y)=$ constant is a solution with energy
$E=-(1+m)/4$ provided $b=f=0,g=1/2,a=t-1/2$ (i.e. $b+f+g=1/2$).  

One can build solutions for higher values of $b+f+g$ from here. In
particular, in case $b+f+g=2M+1/2$, we consider the ansatz
($M=0,1,2,...$)
\be\label{3.22}
Z(y)=\sum_{k=0}^{M} A_k \sn^{2k}
(y)+\cn(y)\dn(y)\sum_{k=0}^{M-1} B_k \sn^{2k} (y)\,,
\ee
while if $b+f+g=2M+3/2$ then we consider the ansatz ($M=0,1,2,...$)
\be\label{3.23}
Z(y)=\cn(y,m)\sum_{k=0}^{M} A_k \sn^{2k}
(y)+\dn(y)\sum_{k=0}^{M} B_k \sn^{2k} (y)\,,
\ee
Substituting this ansatz in eq. (\ref{3.21}) and simplifying, 
one gets analytic expressions for the
energy eigenvalues and eigenfunctions for arbitrary $M$ for 
$b=1/2$ and $b=3/2$. In particular, for $b=1/2$, we find that
\be\label{3.24}
g=1/2\,,~ b+f=N\,,~ a=t-1/2\,,~~E=-[(f+g)^2+m(g+b)^2]\,,
\ee
where both $b,f$ are nonnegative integers satisfying $b+f=N$ with
$N=0,1,2,...$.

Similarly, when $g=3/2,a=t-1/2,b+f=N$ we find that
\be\label{3.25}
E=1+2f+(2b+1)m-[(f+g)^2+m(g+b)^2]\pm \sqrt{(1-m)[(2f+1)^{2}-(2b+1)^{2}
m]
+4mt^2}\,
\ee
where, $b$ and $f$ are again nonnegative integers. In all these cases, the
corresponding eigenfunctions have the form as given above in eqs.
(\ref{3.22}) and (\ref{3.23}). For small  
 values of $N$, the  coefficients $A_k,B_k$ appearing in the
eigenfunctions can be easily written down. 
Further, as in the half-integral $b$ case, one can write down
three more solutions with the same energy. 

\section{\bf Supersymmetry and Potentials with a Finite Number of Band Gaps}
We shall now start with the ground state as well as the excited state
eigenfunctions of various PT-invariant GAL
potentials discussed in the last section and using supersymmetry 
obtain the corresponding SUSY partner potentials. In this manner, we
obtain many new periodic potentials $V_{+}(x)$ with a finite number of band
gaps. As emphasized in the introduction, unlike real potentials, 
if we take a complex PT-invariant potential, then even if we
start with an excited state wave function and calculate the
corresponding superpotential $W$, the
singularities in $W$ and hence $V_{+}(x)$ are not on the real axis, and do not cause problems. 

\subsection{Supersymmetry Partners of PT-Invariant Lam\'e Potentials}

The simplest case is when only one parameter,
(say $a$) is nonzero. This gives the PT-invariant Lam\'e potential 
\be\label{3}
V(x)=-a(a+1)m\sn^2(y)\,.
\ee
For concreteness, take $a=1$, which yields $V(x)=-2m\sn^2(y)$. 
Here, the three band edge
eigenfunctions (in order of increasing energy eigenvalues) are
$\sn(y),\cn(y),$
$\dn(y)$. It is easily computed that
corresponding to these three eigenstates, the corresponding partner
potentials (up to a constant) are $V_{+}(x) =
-2m\sn^2(y+K(m)),-2m\sn^2(y+iK'(m)),
-2m\sn^2(y+K(m)+iK'(m))$ which are all strictly isospectral
potentials to the original Lam\'e potential. Thus, in this case we do
not obtain any new solvable potentials by using supersymmetry.

Now consider the case $a=2$.
All the five band edge
eigenvalues and eigenfunctions of the PT-invariant Lam\'e potential
$V(x)=-6m\sn^2 (y)$ have already been given by us in Table 4 of ref.  
\cite{ks6}. 
Starting from any of the five band edge eigenfunctions and 
calculating the corresponding superpotentials, 
we obtain five different
supersymmetric partner potentials all of which have the same band edge
energy eigenvalues as given in Table 4 of ref. \cite{ks6}. 
In Table 5 we have given the expressions for these
five different strictly isospectral potentials. 
It is worth noting that out of these
five potentials, three are self-isospectral - they are the PT-invariant 
GAL potentials $[2,2,2,0]$. Hence, truly speaking,
we only have three genuinely different potentials, all having the same band 
edge energies. 
For each of these cases, using the
formalism of supersymmetric quantum mechanics \cite{cks}, we can easily obtain
expressions for the corresponding five eigenstates. Now, again by
starting from these eigenfunctions, we can construct still different
partner potentials but with identical band edges. In this way, one
could construct a large number of periodic
potentials with five band edges and two band gaps, all
strictly isospectral to the PT-invariant Lam\'e potential (\ref{3}) with
$a=2$.  

Similarly, if we consider the PT invariant Lam\'e potential (\ref{3}) 
with $a=3$, then we
have 7 band edge eigenfunctions and eigenvalues all of which are analytically known
and are given in Table 1 of ref. \cite{ks5}.
Again, using supersymmetry, we can obtain seven different partner potentials $V_{+}$ 
all with the same band edge eigenvalues. By starting from any
one of them and using other eigenfunctions recursively, we can in
principle construct a huge class of new isospectral potentials. 
Particular mention may be made of the case when we start
from the eigenfunction $\sn(y)\cn(y)\dn(y)$
of the potential $V(x)=-12m\sn^2(y)$. 
It is easily shown
that the corresponding partner potential $V_{+}$ (up to a constant)
is given by
\be\label{8}
V(x)=-m[6\sn^2(y)+2\sn^2(y+K(m))+2\sn^2(y+iK'(m))
+2\sn^2(y+K(m)+iK'(m))]\,.
\ee
Thus, we see that the PT-invariant GAL 
potential $[6,2,2,2]$ 
has precisely three bands, three band gaps and seven
band edges, since it is the supersymmetric partner of the PT-invariant Lam\'e potential
(\ref{3}) with $a=3$. The process described above is readily extended to 
any Lam\'e potential with integer $a$. We
can start from any of the $2a+1$ band edges and obtain the corresponding
supersymmetric partner potentials all having the same band edges. 

We have shown
that the SUSY partners of the PT-invariant Lam\'e potentials $[6,0,0,0]$ and
$[12,0,0,0]$ are the potentials $[2,2,2,0]$ and $[6,2,2,2]$
respectively. What
about the higher Lam\'e potentials? In this connection, it is amusing to
notice that the band edges of the PT-invariant Lam\'e potential $[20,0,0,0]$
and the potential $[6,6,6,2]$ (which follow from Table 4) 
are identical. For example, out of 9 band edges, the 6
band edge energy eigenvalues of $[20,0,0,0]$ are given by
\be\label{3.8}
E=-5(m+2)\pm \sqrt{4m^2-9m+9}\,,~~E=-5(1+m)\pm 2\sqrt{4m^2+m+4}\,,~~
E=5(1+2m)\pm 2\sqrt{9m^2-9m+4}\,.
\ee
It is easily seen from Table 4 that exactly the 
same eigenvalues are obtained when $a,b,f,g$ take the values
$(2,2,-3,1),(2,-3,2,1),(-3,2,2,1)$. Similarly, one can show that the
three remaining eigenvalues of $[20,0,0,0]$ satisfy the same cubic equation 
as $[6,6,6,2]$ when $a,b,f,g$ take the values $(2,2,2,-2)$. 

In fact, one can show that the number (and structure) of band edges of the
PT-invariant Lam\'e potential $[2a(2a+1),0,0,0]$ is same as the QES states
of the potential $[a(a+1),a(a+1),a(a+1),(a-1)a]$. For 
example, for this PT-invariant Lam\'e potential 
it is well known that out of the $4a+1$ band edges
of the Lam\'e potential, $a$ states each are of the form
$\cn(y)\sn(y) F_{a-1}(\sn^2(y))$,
$\cn(y)\dn(y) F_{a-1}(\sn^2(y))$,
$\dn(y)\sn(y) F_{a-1}(\sn^2(y))$, while the remaining $a+1$ states are 
of the form 
$F_{a}(\sn^2(y))$. Using Table 4, it is
easily shown that there are again $4a+1$ QES states of 
the potential
$[a(a+1),a(a+1),a(a+1),(a-1)a]$, out of which $a$ QES states each are
obtained when $a,b,f,g$ are of the form $a,a,a-1,a-1$, 
or $a,a-1,a,a-1$,
or $a-1,a,a,a-1$, while $a+1$ QES states are obtained when $a,b,f,g$ are
of the form $a,a,a,-a$. In fact we believe that all the band edge
eigenvalues of the potentials $[2a(2a+1),0,0,0]$ and
$[a(a+1),a(a+1),a(a+1),(a-1)a]$ are identical. While this is easily
shown for low values of $a$, at the moment, a general proof is still lacking.

Similarly, one can show that the number (as well as the structure) of 
band edges of the PT-invariant 
Lam\'e potential $[(2a-1)2a,0,0,0]$ is the same as the QES states
of the potential $[a(a+1),(a-1)a,(a-1)a,(a-1)a]$. For
example, it is well known that out of the $4a-1$ band edges
of the PT-invariant Lam\'e potential, $a$ states each are of the form
$\cn(y) F_{a-1}(\sn^2(y))$,
$\dn(y) F_{a-1}(\sn^2(y))$,
$\sn(y) F_{a-1}(\sn^2(y))$, while the remaining $a-1$ states are 
of the form 
$ \sn(y)\cn(y)\dn(y)$$F_{a-2}(\sn^2(y)$. 
Using Table 4, it is
easily shown that there are $4a-1$ QES states of 
the potential
$[a(a+1),(a-1)a,(a-1)a,(a-1)a]$, out of which $a$ QES states each are
obtained when $a,b,f,g$ are of the form $a,-a,a-1,a-1$, or $a,a-1,-a,a-1$,
or $a,a-1,a-1,-a$, while $a-1$ QES states are obtained when $a,b,f,g$ are
of the form $-a-1,a-1,a-1,a-1$. 
In fact we believe that all the band edge
eigenvalues of the potentials $[(2a-1)2a,0,0,0]$ and 
$[a(a+1),(a-1)a,(a-1)a,(a-1)a]$ are identical. While this is easily
shown for low values of $a$, a general proof is not available.

On the basis of these results, we then conjecture that the potentials 
$[a(a+1),a(a+1),a(a+1),(a-1)a]$, for integer $a$,
 have the same band edges
as the Lam\'e potential $[2a(2a+1),0,0,0]$ and hence these potentials
also have precisely $2a$ band gaps and $(4a+1)$ band edges, all of
which are known in principle. Further, the potentials 
$[a(a+1),(a-1)a,(a-1)a,(a-1)a]$ have the same band edges as the 
Lam\'e potential $[(2a-1)2a,0,0,0]$ and hence are also potentials with a 
finite number ($2a-1$) of band gaps. It would be nice to have a general proof.

\subsection{\bf Supersymmetry Partners of PT-Invariant 
Associated Lam\'e Potentials}

We start our discussion with the $a=2$, $b=1$
associated Lam\'e potential and its corresponding PT-invariant potential 
$V^{PT}(x)=-6m\sn^2(y)-2m{\cn^2 (y)}/{\dn^2(y)}\,$.  All 
five band-edge eigenvalues and eigenfunctions for this potential 
have been given by us in
Table 3 of ref. \cite{ks6}. 
As established previously \cite{ks1,ks6}, this is a self-isospectral potential and
hence using the  band edge eigenfunction $\dn^2(y)$ 
does not give any new partner
potential. However, if instead we use the remaining four band edge
eigenfunctions, then one gets four new SUSY partner potentials 
which are strictly isospectral to the PT-invariant [6,2,0,0]
potential.

Let us now consider the PT-invariant AL potential
$[a(a+1), (a-2)(a-1),0,0]$, i.e. the potential (\ref{2}) with $b=a-2,
f=g=0$. 
As shown by us
\cite{ks1}, one of its exact band edge eigenfunction is 
$\psi(x) = \cn(y)\dn^{a-1} (y)$. It is easy to see that
the corresponding partner potential $V_{+}$ (up to a constant) is
the potential  $[(a-1)a,(a-1)a,2,0]$.
Thus we immediately conclude that the PT-invariant potential 
$[(a-1)a,(a-1)a,2,0]$ 
is strictly isospectral to
the PT-invariant AL potential $[a(a+1),(a-2)(a-1),0,0]$. In the 
special case when both $a,b$ are integers, in view of our results on
AL potentials \cite{ks2}, it then follows that the
GAL potential $[(a-1)a,(a-1)a,2,0]$   
has $a$ band gaps and $a$ bands, out of which  $b=a-2$ 
bands are rather unusual. 

Note that if instead we use
$\psi(x) = \sn(y)\dn^{a-1} (y,m)$, which is also one of the exact
eigenfunctions of the above AL potential, then nothing new is obtained. 
In particular, the corresponding partner potential is
$[(a-1)a,(a-1)a,0,2]$ which is strictly isospectral to the potential
$[(a-1)a,(a-1)a,2,0]$.

Let us now consider the PT-invariant AL potential
$[a(a+1), (a-3)(a-2),0,0]$, i.e. the potential (\ref{2}) with $b=a-3,
f=g=0$. 
As shown by us
\cite{ks1}, one of its exact band edge eigenfunction is 
$\psi(x) = \sn(y) \cn(y)\dn^{a-2} (y)$. It is easy to see that
the corresponding partner potential $V_{+}$ (up to a constant) is
the potential $[(a-1)a,(a-2)(a-1),2,2]$ in the notation of (\ref{2}).
Thus we immediately conclude that when $a,b$ are integers, then 
this PT-invariant potential 
is strictly isospectral to 
the AL potential $[a(a+1),(a-3)(a-2),0,0]$, has $a$ band gaps and $a$
bands, out of which $b=a-3$ bands are rather unusual. 

We can generalize the above arguments. In particular, we find that the number 
(and even
structure) of the potentials
$[(a-p)(a-p+1),(a-p-1)(a-p),p(p+1),p(p+1)]$ is the same as the AL
potentials $[a(a+1),(a-2p-1)(a-2p),0,0]$. For example, as remarked
in the previous section, if 
$b=a-2p-1 (p=0,1,2,...$), then there are $p$ eigenstates of the form
$\sn(y)\cn(y)\dn(y)$$F_{p-1}(\sn^2(y))$, $p+1$ eigenstates of
the form
$\dn^{a-2p}(y)$$F_{p}(\sn^2(y))$, $a-p$ eigenstates of the form
$\cn(y)\dn^{2p+1-a}(y)$$F_{a-p-1}(\sn^2(y))$ and also $a-p$ eigenstates 
of the form
$\sn(y)\dn^{2p+1-a}(y)$$F_{a-p-1}(\sn^2(y))$. Using Table 4
it is easy to show that for
the GAL potential,
$[(a-p)(a-p+1),(a-p-1)(a-p),p(p+1),p(p+1)]$, there are
$p$ eigenstates of the form
$\sn^{-p}(y)$$\cn^{-p}(y)$
$\dn^{1+p-a}(y)$$F_{p-1}(\sn^2(y))$, 
$p+1$ eigenstates of
the form
$\dn^{a-p}(y)$$\cn^{-p}(y)$
$\sn^{-p}(y)$$F_{p} (\sn^2(y))$, 
$a-p$ eigenstates of the form
$\cn^{p+1}(y)$$\sn^{-p}(y)$
$\dn^{p+1-a}(y)$
$F_{a-p-1}(\sn^2(y))$ 
and also $a-p$ eigenstates 
of the form
$\sn^{p+1}(y)$$\cn^{-p}(y)$
$\dn^{p+1-a}(y)$$F_{a-p-1}(\sn^2(y))$. 
In fact we believe that all the band edge
eigenvalues of the potentials $[a(a+1),(a-2p)(a-2p+1),0,0]$ and
$[(a-p)(a-p+1),(a-p-1)(a-p),p(p+1),p(p+1)]$ are identical. While this is easily
shown for low values of $a$ and $p$, a general proof is still lacking.

Similarly, we can show that the number (and even
structure) of the potentials
$[(a-p)(a-p+1),(a-p)(a-p+1),p(p+1),(p-1)p]$ is the same as the AL
potentials $[a(a+1),(a-2p)(a-2p+1),0,0]$. 
In particular, for the AL potential, as shown in Sec. 2,
when $b=a-2p$, then there are $p$ eigenstates of the
form 
$\cn(x)\dn^{a-2p+1}(x)$$F_{p-1}(\sn^2(x))$, $p$ eigenstates 
of the form
$\sn(x)$$\dn^{a-2p+1}(x)$$F_{p-1}(\sn^2(x))$, $a-p$ eigenstates
of the form
$\sn(x)$$\cn(x)\dn^{2p-a}(x)$$F_{a-p-1}(\sn^2(x))$, and $a-p+1$ 
eigenstates of the form
$\dn^{2p-a}(x)F_{a-p}(\sn^2(x))$.
It is easily shown that for the potential 
$[(a-p)(a-p+1),(a-p)(a-p+1),p(p+1),(p-1)p]$, there are $4a-1$ QES
states of similar form. In particular, there are
$p$ eigenstates of the form
$\sn^{-p}(y)\cn^{1-p}(y)$
$\dn^{p-a}(y)$$F_{p-1}(\sn^2(y))$, 
$p$ eigenstates of
the form
$\dn^{a+1-p}(y)\cn^{1-p}(y)$
$\sn^{-p}(y)$$F_{p-1}(\sn^2(y))$, 
$a-p+1$ eigenstates of the form
$\cn^{p}(y)$$\sn^{-p}(y)$
$\dn^{p-a}(y)$$F_{a-p}(\sn^2(y))$ 
and also $a-p$ eigenstates 
of the form
$\sn^{p+1}(y)$$\cn^{1-p}(y)$
$\dn^{p-a}(y)$
$F_{a-p-1}(\sn^2(y))$. 
In fact we believe that all the band edge
eigenvalues of the potentials $[a(a+1),(a-2p)(a-2p+1),0,0]$ and
$[(a-p)(a-p+1),(a-p)(a-p+1),p(p+1),(p-1)p]$ are identical. While this is easily
shown for low values of $a$ and $p$, we don't yet have a general proof.

\subsection{\bf SUSY Partners of Potentials with $b=f=0$}

Let us now consider the SUSY partners of the potential
$[a(a+1),0,0,g(g+1)]$  which for integral values of $a,g$, is a problem with a  
finite number of band gaps. By exactly following the above discussion about the
PT-invariant AL potential, we can construct a
host of new potentials with a finite number of band gaps. For example, 
by starting from the potential $[6,0,0,2]$ and following the procedure
as in the AL case, we can
easily obtain four new SUSY partner potentials, all with two band
gaps. 

From Table 1 we observe that for integral $a$, 
two of the exact eigenfunctions of
the potential $[a(a+1),0,0,(a-2)(a-1)]$ with $a$ band gaps
 are $\cn(y)\sn^{a-2}(y)$
and $\dn(y)\sn^{a-2}(y)$. It is easily seen that if we start with
either of these eigenfunctions, then the corresponding SUSY partner
potential with the same finite ($a$) number of band gaps is the
potential $[(a-1)a,2,,0,(a-1)a]$ (or its isospectral
partner $[(a-1)a,0,2,(a-1)a]$). 

From Table 1 we also observe that one of the exact eigenfunction of
the potential $[a(a+1),0,0,(a-3)(a-2)]$ is
$\cn(y,m)\dn(y,m)\sn^{a-3}(y,m)$. On starting with this eigenfunction,
it is easily shown that the corresponding SUSY partner potential is
$[(a-1)a,2,2,(a-2)(a-1)]$ which therefore must also be a potential
with finite ($a$) number of band-gaps in case $a$ is an integer.

Similarly, by starting from the finite band-gap 
potentials $[a(a+1),0,0,(a-2p-1)(a-2p)]$ as well as
$[a(a+1),0,0,(a-2p)(a-2p+1)]$, and following the discussion in the
case of PT-invariant AL potential, it is easily shown that the
corresponding SUSY partners with the same (finite) number of band gaps
are the potentials
$[(a-p)(a-p+1),p(p+1),p(p+1),(a-p-1)(a-p)]$ and
$[(a-p)(a-p+1),p(p+1),(p-1)p,(a-p)(a-p+1)]$ and
respectively, where $a$ and $p$ are positive integers.

\subsection{\bf SUSY Partners of Potentials with $b=g=0$}

Let us now consider the SUSY partners of the PT-invariant potential
$[a(a+1),0,f(f+1),0]$. 
{}From Table 2 we observe that two of the exact eigenfunctions of
the potential $[a(a+1),0,(a-2)(a-1),0]$ are $\sn(y)\cn^{a-2}(y)$
and $\dn(y)\cn^{a-2}(y)$. It is easily seen that if we start with
either of these eigenfunctions, then the corresponding SUSY partner
potentials turn out to be $[(a-1)a,2,(a-1)a,0]$ or
$[(a-1)a,0,(a-1)a,2]$. Since we know that the potentials 
$[a(a+1),a(a-1),2,0]$ as well as $[a(a+1),2,0,(a-1)a]$ have a finite number of band gaps,  
we conjecture that maybe the potential $[a(a+1),2,(a-1)a,0]$ 
also has only a finite number ($a$) of band gaps when 
$a$ is an integer.

{}From Table 2 we also observe that one of the exact eigenfunctions of
the potential $[a(a+1),0,(a-3)(a-2),0]$ is
$\sn(y)\dn(y)\cn^{a-3}(y)$. Starting with this eigenfunction,
it is easily shown that the corresponding SUSY partner potential is
$[(a-1)a,2,(a-2)(a-1),2]$. Again, since for integer $a$, the potential 
$[(a-1)a,(a-2)(a-1),2,2]$ has only a finite number of band gaps, it is
tempting to conjecture that the same may also be true for the
potential $[(a-1)a,2,(a-2)(a-1),2]$. 

Similarly, by starting from the finite band-gap 
potentials $[a(a+1),0,(a-2p-1)(a-2p),0]$ as well as
$[a(a+1),0,(a-2p)(a-2p+1),0]$, and following the discussion in the
case of PT-invariant AL potentials, it is easily shown that the
corresponding SUSY partners with the same number of band gaps
are the GAL potentials
$[(a-p)(a-p+1),p(p+1),p(p+1),(a-p-1)(a-p)]$ and
$[(a-p)(a-p+1),p(p+1),(p-1)p,(a-p)(a-p+1)]$
respectively when $a$ and $p$ are integers.

\subsection{\bf SUSY Partners of Potentials with $f=0$}

Let us now consider the SUSY partners of the potential
$[a(a+1),b(b+1),0,g(g+1)]$.
{}From Table 3 we observe that one of the exact eigenfunctions 
is $\dn^{-b}(y)\sn^{-g}(y)$
when $a+b+g=0$. If we start with
this eigenfunction, then the corresponding SUSY partner
potential turns out to be $[(a-1)a,(b-1)b,0,(g-1)g]$.

{}From Table 3 we also observe that an exact eigenfunction of
the potential $[a(a+1),b(b+1),0,g(g+1)]$ is
$\cn(y)\dn^{-b}(y)\sn^{-g}(y)$ when $a+b+g=1$. 
Starting with this eigenfunction,
it is easily shown that the corresponding SUSY partner potential is
$[(a-1)a,(b-1)b,2,(g-1)g]$. 

In summary, we have discovered a large number of complex PT-invariant 
periodic potentials with a finite number of band gaps, many occurring 
when the parameters $a,b,c,d$ have specific
integer values. This leads us to make the plausible conjecture 
that all GAL potentials (\ref{2}) for integer values of
$a,b,f,g$ have a
finite number of band-gaps, but there is as yet no formal proof.

\section{Heun's Equation and the Generalized Associated Lam\'e Equation}
In this section, we point out an interesting connection between 
Heun's differential equation \cite{ren} and the generalized associated Lam\'e
equation (\ref{3.2}). This connection enables us to use the various solutions 
of eq. (\ref{3.2}) obtained in this paper to write down 
several solutions of Heun's equation which have apparently not been studied in the mathematics literature.

The canonical form of Heun's equation is given by \cite{ren}
\be\label{4.1}
\bigg [\frac{d^2}{dx^2}+\big (\frac{\gamma}{x}+\frac{\delta}{x-1}
+\frac{\epsilon}{x-c} \big )\frac{d}{dx}+\frac{\alpha \beta
x-q}{x(x-1)(x-c)} \bigg ]G(x) =0\,,
\ee
where $\alpha,\beta,\gamma,\delta,\epsilon,q,c$ are real parameters, except
that $c \ne 0,1$ and the first five parameters are constrained by the
relation
\be\label{4.1a}
\gamma+\delta+\epsilon=\alpha+\beta+1\,.
\ee
If we make the transformation $x=\sn^2(y,m)$, then Heun's equation takes
the form \cite{ren}
\bea\label{4.2}
&&F''(y)+[(1-2\epsilon) m \frac{\sn(y) \cn(y)}{\dn(y)}
+(1-2\delta)
\frac{\sn(y)\dn(y)}{\cn(y)} +(2\gamma-1) \frac{\cn(y)
\dn(y)}{\sn(y)}]F'(y) \nonumber \\
&&-[4mq
-4\alpha \beta m \sn^2 (y)]F(y) =0\,,
\eea
where $[G(x) \equiv F(y)]$ and  $m=1/c$.
It is interesting to note that eq. (\ref{4.2}) 
is very similar to  the $\phi$ equation
(\ref{3.4}) which we have analyzed in great detail.
In particular, with the identification 
\be\label{4.3}
b=\frac{1}{2}-\epsilon\,,~f=\frac{1}{2}-\delta\,,~g=\frac{1}{2}-\gamma\,,
~b+f+g=\frac{1}{2}-\alpha-\beta, 4\alpha \beta = Q\,,
4mq=R\,,
\ee
all the results discussed above 
can be immediately
used to obtain different solutions of Heun's equation. 
It turns out that using the mid-band states obtained in 
Sec. 2, one
generates new quasi-periodic solutions of Heun's eq. (\ref{4.2}),
which we discuss in a separate publication \cite{ks7}.

\newpage


\noindent{\bf Table 1:} Energy eigenstates of PT-invariant 
GAL potentials with $b=f=0,$
$g=n-a,$ $n=0,1,2,\ldots$;  
$\delta_1 \equiv \sqrt{(1+m)^2(a-1)^2-(2a-1)(2a-3)m}\,,~\delta_2 \equiv  
\sqrt{[a-1+m(a-2)]^2
-(2a-1)(2a-5)m}\,,\\
\delta_3 \equiv \sqrt{[a-2+m(a-1)]^2-(2a-1)(2a-5)m}\,,~\delta_4 \equiv 
\sqrt{(1+m)^2(a-2)^2 -(2a-1)(2a-7)m}$\,. \sss

\oddsidemargin      -0.2in

\bigskip
\begin{tabular}{cccccc}
\hline
$n$ & $g(g+1)$ &  $E$ & $\sn^{-a} (y) \psi$\\
\hline
$0$ & $(a-1)a$&  $-(1+m)a^2$ & $1$\\
$1$ & $(a-2)(a-1)$&  $ -a^2-m(a-1)^2$ & 
$\frac{\cn (y)}{\sn(y)}$\\
$1$ & $(a-2)(a-1)$ &  $ -(a-1)^2-ma^2$ & 
$\frac{\dn (y)}{\sn(y)}$\\
$2$ & $(a-3)(a-2)$ &  $ -(1+m)(a-1)^2$ & 
$\frac{\cn (y) \dn(y)}{\sn^2(y)}$\\
$2$ & $(a-3)(a-2)$ &  $ -(1+m)(a^2-2a+2) \pm 2\delta_1$ & 
$\frac{[E+(1+m)(a-2)^2]\sn^2(y)+2(2a-3)}{\sn^2(y)}$\\
$3$ & $(a-4)(a-3)$ &  $ -(a^2-2a+2)-(a^2-4a+5)m \pm 2\delta_2$ & 
$\frac{[[E+(a-2)^2+m(a-3)^2]\sn^2(y)+2(2a-5)]\cn(y)}{\sn^3(y)}$\\
$3$ & $(a-4)(a-3)$  & $ -(a^2-4a+5)-(a^2-2a+2)m \pm 2\delta_3$ & 
$\frac{[[E+(a-3)^2+m(a-2)^2]\sn^2(y)+2(2a-5)]\dn(y)}{\sn^3(y)}$\\
$4$ & $(a-5)(a-4)$ &  $ -(1+m)(a^2-4a+5) \pm 2\delta_4$ & 
$\frac{[[E+(1+m)(a-3)^2]\sn^2(y)+2(2a-7)]\cn(y)\dn(y)}{\sn^4(y)}$\\
\hline
\end{tabular}
\bigskip

\vskip 0.9 true cm
\noindent{\bf Table 2:} Energy eigenstates of PT-invariant 
GAL potentials with parameters $b=g=0,$ 
$f=n-a,$ $n=0,1,2,\ldots$; 
 $\delta_5 \equiv \sqrt{(a-1+m)^2-(2a-1)m}\,,~\delta_6 \equiv \sqrt{(a-1+2m)^2
-3(2a-1)m}\,,\\
\delta_7 \equiv \sqrt{(a-2+2m)^2-(2a-1)m}\,,~\delta_8 \equiv
\sqrt{(a-2+3m)^2-3(2a-1)m}$\,. \sss
\oddsidemargin      -0.2in

\bigskip
\begin{tabular}{cccc}
\hline
$n$ & $f(f+1)$ & $E$ & $\cn^{-a} (y) \psi$   
 \\
\hline
$0$ & $(a-1)a$ &  $-a^2$ & $1$\\
$1$ & $(a-2)(a-1)$ &  $ -a^2-m$ & 
$\frac{\sn (y)}{\cn(y)}$\\
$1$ & $(a-2)(a-1)$  &  $ -(a-1)^2-m$ & 
$\frac{\dn (y)}{\cn(y)}$\\
$2$ & $(a-3)(a-2)$ &  $ -(a-1)^2-4m$ & 
$\frac{\dn (y) \sn(y)}{\cn^2(y)}$\\
$2$ & $(a-3)(a-2)$ &  $ -(a^2+2-2a+2m) \pm 2\delta_5$ & 
$\frac{(E+(a-2)^2)\sn^2(y)+2}{\cn^2(y)}$\\
$3$ & $(a-4)(a-3)$ &  $ -(a^2+2-2a+5m) \pm 2\delta_6$ & 
$\frac{[(E+(a-2)^2+m)\sn^2(y)+6]\sn(y)}{\cn^3(y)}$\\
$3$ & $(a-4)(a-3)$& $ -(a^2+5-4a+5m) \pm 2\delta_7$ & 
$\frac{[(E+(a-3)^2+m)\sn^2(y)+2]\dn(y)}{\cn^3(y)}$\\
$4$ & $(a-5)(a-4)$ & $ -(a^2+5-4a+10m) \pm 2\delta_8$ & 
$\frac{[(E+(a-3)^2+4m)\sn^2(y)+6]\dn(y)\sn(y)}{\cn^4(y)}$\\
\hline
\end{tabular}
\bigskip

\vskip 0.8 true cm
\newpage

\noindent{\bf Table 3:} Energy eigenstates of PT-invariant 
GAL potentials with 
$f=0$, $g=n-a-b,$  $n=0,1,2,\ldots$; 
$\delta_9 \equiv \sqrt{[(1+m)(a-1)+b]^2-(2a-1)(2a+2b-3)m},\\
\delta_{10} \equiv  
\sqrt{[a+b-1+m(a-2)]^2-(2a-1)(2a+2b-5)m}$. \sss

\oddsidemargin      -0.2in

\bigskip
\begin{tabular}{cccccc}
\hline
$n$ & $E$ & $\dn^{b} (y) \sn^{-(a+b)} (y)\psi$\\
\hline
$0$ &  $-(a+b)^2-ma^2$ & $1$\\
$1$ &  $ -(a+b)^2-m(a-1)^2$ & 
$\frac{\cn (y)}{\sn(y)}$\\
$2$ &  $ -(1+m)-(a+b-1)^2-m(a-1)^2 \pm 2\delta_9$ & 
$\frac{[E+(a+b-2)^2+m(a-2)^2]\sn^2(y)+2(2a+2b-3)}{\sn^2(y)}$\\
$3$ &  $ -(1+m)-(a+b-1)^2-m(a-2)^2 \pm 2\delta_{10}$ & 
$\frac{[[E+(a+b-2)^2+m(a-3)^2]\sn^2(y)+2(2a+2b-5)]\cn(y)}{\sn^3(y)}$\\
\hline
\end{tabular}

\vskip 1.8 true cm
\noindent{\bf Table 4:} Energy eigenstates of PT-invariant 
GAL potentials with 
$f=2n-a-b-g,$  $n=0,1,2,\ldots$; 
$\delta_{11} \equiv \sqrt{[(a+b-1)+m(1-b-g)]^2-(2a-1)(1-2g)m}$\,. \sss

\oddsidemargin      -0.2in

\bigskip
\begin{tabular}{cccccc}
\hline
$n$ &  $E$ & $\sn^{g} (y) \dn^{b} (y) \cn^{-(a+b+g)} (y) \psi$\\
\hline
$0$ &  $-(a+b)^2-m(g+b)^2$ & $1$\\
$1$ &  $ -(a+b-1)^2-m(b+g-1)^2-(1+m)\pm 2\delta_{11}$ & 
$\frac{[E+(a+b-2)^2+m(b+g)^2]\sn^2(y)-2(2g-1)}{\cn^2(y)}$\\
\hline
\end{tabular}

\vskip 1.8 true cm

\noindent {\bf Table 5:} The five supersymmetric partner potentials 
of the PT-invariant Lam\'e potential $V_{-}^{PT}(x) =
-6m\sn^2(y)$.
Here $y=ix+\beta$ and $\delta \equiv \sqrt{1-m+m^2}$. 
All partner potentials have a period $2K'(m).$ \sss 

\bigskip
\begin{tabular}{cccccc}
\hline
 $E$ & $\psi^{(-)}$ & 
$V_{+}(x)$\\
\hline
 $-2(1+m)-2\delta$ & $1+\frac{E}{2}\sn^2 (y)$ &
$6m\sn^2(y)+E-\frac{2E^2\sn^2(y)\cn^2(y)\dn^2(y)}{(1+\frac{E}{2}\sn^2(y))}       $\\
 $-4-m$ & $\sn(y)\cn(y)$ &
$-2m[\sn^2(y)+\sn^2(y+K(m)+iK'(m))+\sn^2(y+iK'(m)] -E$\\
 $-1-4m$ & $\sn(y)\dn(y)$ &
$-2m[\sn^2(y)+\sn^2(y+K(m))+\sn^2(y+iK'(m)] -E$\\
 $-1-m$ & $\cn(y)\dn(y)$ &
$-2m[\sn^2(y)+\sn^2(y+K(m))+\sn^2(y+K(m)+iK'(m)] -E$\\
 $-2(1+m)+2\delta$ & $1+\frac{E}{2}\sn^2 (y)$ &
$6m\sn^2(y)+E-\frac{2E^2\sn^2(y)\cn^2(y)\dn^2(y)}{(1+\frac{E}{2}\sn^2(y))}       $\\
\hline
\end{tabular}
\bigskip
\end{document}